\documentclass[fleqn,10pt]{wlscirep}
\usepackage[utf8]{inputenc}
\usepackage[T1]{fontenc}
\usepackage{tcolorbox}
\usepackage{caption}
\usepackage{subcaption}
\usepackage{makecell}
\usepackage{mciteplus}
\usepackage{hyperref}
\usepackage{sidecap}
\sidecaptionvpos{figure}{t}
\usepackage{amssymb,amsmath}
\usepackage{float}
\usepackage{lineno}
\usepackage{color}

\definecolor{niceblue}{RGB}{0, 91, 187}

\hypersetup{
    colorlinks=true,
    linkcolor=niceblue,
    filecolor=magenta,      
    urlcolor=niceblue,
    pdftitle={Defining and classifying models of groups},
    }

\title{Defining and classifying models of groups: The social ontology of higher-order networks}

\author[1]{Jonathan St-Onge}
\author[2]{Randall Harp}
\author[1,3]{Giulio Burgio}
\author[4]{Timothy M. Waring}
\author[1,4,6]{Juniper Lovato}
\author[1,4,6,7]{Laurent H\'ebert-Dufresne}
\affil[1]{Vermont Complex Systems Institute, University of Vermont, Burlington VT, USA 05405}
\affil[2]{Department of Philosophy, University of Vermont, Burlington VT, USA 05405}
\affil[3]{Institute of Functional Biology and Genomics IBFG-CSIC, University of Salamanca, 37007 Salamanca, Spain}
\affil[4]{Department of Computer Science, University of Vermont, Burlington VT, USA 05405}
\affil[5]{School of Economics and Mitchell Center for Sustainability Solutions, University of Maine, Orono ME, USA 04469}
\affil[6]{Complexity Science Hub, A-1080 Vienna, Austria}
\affil[7]{Santa Fe Institute, Santa Fe NM, USA 87501}

\begin{abstract}

In complex systems research, the study of higher-order interactions has exploded in recent years. Researchers have formalized various types of group interactions, such as public goods games, biological contagion, and information broadcasting, showing how higher-order networks can capture group effects more directly than pairwise models. However, equating hyperedges---edges involving more than two agents---with groups can be misleading, as it obscures the polysemous nature of the term ``group interactions''. For instance, many models of higher-order interactions focus on the internal state of the hyperedge, specifying dynamical rules at the group level. In doing so, these models often neglect how interactions with external groups can influence behaviors and dynamics within the group itself. Yet, anthropologists and philosophers remind us that external norms, factors, and forces governing intergroup behavior, in the form of intergroup competition or cooperation, are essential to defining within-group dynamics. In this paper, we synthesize concepts from social ontology relevant to the emerging physics of higher-order networks. We propose a typology for classifying models of group interactions based on two key perspectives. The first focuses on individuals within groups engaging in collective action, where shared agency serves as the binding force. The second adopts a group-first approach, emphasizing institutional facts that extend beyond the specific individuals involved. Building on these perspectives, we introduce four dimensions to classify models of group interactions: persistence, coupling, reducibility, and alignment. For the physics of higher-order networks, we provide a hierarchy of nested mathematical models to explore the complex properties of social groups. We also highlight social interactions that have not yet been explored in the literature on higher-order networks and propose future research avenues to foster collaboration between social ontology and the physics of complex systems.

\end{abstract}
\begin{document}

\flushbottom
\maketitle

\thispagestyle{empty}

\section{Introduction}\label{section:intro}

\noindent Broadly defined, the mathematical modeling of complex systems \cite{hebert-dufresne_path_2024}, and in particular network science \cite{newman_networks_2010}, is the multidisciplinary study of how interactions in a system shape the emergent properties and functions of the system.
In recent years, there has been a large wave of research on higher-order networks where group interactions are modeled more directly as hyperedges, which can be different entities based on their \textit{order} and size \cite{battiston_physics_2021}. With hyperedges, group interactions at a higher order, such as adolescent cliques, committed minorities, information broadcasting, or contagion within households, can be shown to be \textit{different} from the pairwise interactions at a lower order \cite{ferraz_de_arruda_contagion_2024}. By modeling group interactions at higher orders, we can analyze them as more than mere correlated bundles of pairwise interactions. However, human group interactions extend beyond individuals interacting with one another. Group norms and institutions shape group behaviors just as groups influence individual dynamics, yet they remain irreducible to multi-way interactions between individuals. For instance, mask mandates and other shielding measures significantly alter contagion dynamics in workplaces, extending beyond the effects of higher-order interactions alone \cite{st-onge_paradoxes_2024}. These public health policies are the end result of cumulative cultural learning, shaped by thousands of years of socio-technical innovations and intergroup competition \cite{boyd_culture_1988, tomasello_why_2009, henrich_weirdest_2020}. Similarly, while an increase in the number of coauthors in papers may hint at the rise of teams in science \cite{wuchty_increasing_2007, uzzi_scientific_2012}, modeling the norms and practices within teams directly offers a different perspective on the role of collaboration in shaping scientific productivity.

To model networks beyond higher-order interactions, we propose a typology of group interactions that is informed by the long-standing history of ``group minds'' in the social sciences and, more broadly, the social ontology of groups. We use group minds to refer to a recurring idea across disciplines and time periods---the idea that societies and human groups can exhibit emergent, group-level properties. We establish two key connections. The first highlights the impact of group interactions on individuals, particularly as influenced by varying degrees of group persistence and coupling. Just as the Condorcet's voting paradox and other anti-aggregation arguments in social choice theory demonstrate that group attitudes can contradict individual preferences \cite{list_aggregating_2004}, higher-order interactions display nonlinear behavior as they cannot be reduced to a simple additive function of independent individual influences. 

The second connection focuses on strongly non-reducible groups, distinguishing them from groups that are weakly emergent due to the absence of a simple aggregation scheme. Whereas weak emergence can simply arise from the complex organization of group members, we define strongly non-reducible groups as those with norms and institutions that do not directly follow from the states of their members. By adopting a group-level perspective that emphasizes the role of institutional facts, we argue that we can better investigate the coevolution of individuals and groups---an aspect that is difficult to capture by focusing on individual-based dynamics. We can then show how possible dissonance between individuals and institutions can emerge by reintroducing individual preferences within the context of group dynamics based on group-level fitness. This approach enables us to model group epistemology, cognition, and rationality within higher-order network models, providing a framework for understanding the norms and institutions that govern group interactions.

Taken together, we establish a typology of group models with the following dimensions: persistence, coupling, reducibility, and alignment. This typology helps address ongoing disagreements about the underlying ontology of models of groups. Consider the fact that reviews of higher-order networks often focus on group effects in dynamical processes and group structure in network data, without ever formally defining groups or their properties \cite{ferraz_de_arruda_contagion_2024}. Consequently, it becomes difficult to know whether different modeling techniques agree on group definitions or whether they account for equivalent group features or not. For instance, complex contagion dynamics based on social reinforcement are often not considered ``higher-order'', even when they occur over groups \cite{watts_influentials_2007, osullivan_mathematical_2015}. Colloquially, one key distinction for a system to be considered a higher-order network model appears to be whether the dynamical rules are nonlinear and independent of interactions outside the group.

For example, in our typology, we will distinguish between models assuming isolation and those depending on the states of agents outside the group, or their \textit{coupling}. If coupled, the dynamics within a hyperedge are not fully determined by the state of that hyperedge, and such complex dynamics might not fall under the purview of the current higher-order network literature \cite{ferraz_de_arruda_contagion_2024}. Interestingly, this distinction is more rooted in mathematical convenience than in empirical evidence \cite{ugander_structural_2012, burgio_characteristic_2025}. Conversely, in many evolutionary models of groups, the role of other groups is essential to understand focal groups \cite{boyd_culture_1988, smaldino_evolution_2018}. Our typology is meant to help compare these different definitions and modeling frameworks for groups.

The rest of the paper proceeds as follows. We first provide a brief history of group minds, highlighting the debate surrounding group realism in the sciences (Sec. \ref{interface.section.review}). By examining how beliefs about the existence of groups have shaped social science methodologies, we can better address inconsistencies about the nature of groups in contemporary models of group interactions. Next, we develop a hierarchy of nested mathematical models to explore increasingly complex and detailed properties of social groups (Sec. \ref{typology.intro}). We begin by presenting models of group persistence and coupling mapped to group interactions derived from aggregation-based arguments. We then introduce models of group irreducibility, incorporating the concept of individual-group misalignment. In its most detailed form, our framework allows us to mathematically consider important questions such as collective beliefs, shared intentions, and emerging institutions.  

\section{A short history of group minds}\label{interface.section.review}

Understanding how group behavior emerges from individuals, and when it cannot be reduced to them, has long challenged both social scientists and modelers. In this section, we examine foundational debates over the ontology of groups, particularly the tension between methodological individualism and group realism, as they have shaped how researchers conceptualize and model group interactions. Our aim is to show how differing ontological commitments continue to inform and constrain contemporary modeling choices. By foregrounding these debates, we motivate the conceptual axes of our typology and clarify the assumptions embedded in different approaches to modeling higher-order interactions. This review thus serves to situate our framework within a long-standing intellectual tradition, while also highlighting where existing models may oversimplify or obscure key aspects of social group dynamics.

\paragraph*{The early debate} A longstanding question in the social sciences concerns whether groups are distinct entities with their own dynamics or just convenient labels for aggregates of individuals. This debate is best understood as one between group realism and methodological individualism (MI), shaped by early theoretical developments in sociology and economics \cite{schumpeter_concept_1909, weber_categories_1913}. Group realists like Durkheim emphasized the emergent properties of society, arguing that social norms and institutions regulate behavior and exist independently of individual intentions. While this may seem like a strong ontological commitment---one that prompted disagreements then as now---consider how professions or legal codes can hardly be reduced to any single individual \cite[p.51]{durkheim_rules_1895}. And yet, paradoxically, they are enacted by individuals all the same. MI, as championed by Weber and Schumpeter, argued that only individuals have desires and beliefs; thus, social phenomena must ultimately be explained in terms of individual motives and actions. Schumpeter coined the term ``methodological individualism'' to clarify that treating groups as agents is not merely misleading---it is a category error that confuses individual-level properties with collective entities \cite{schumpeter_concept_1909}. Yet, the issue persisted in different forms throughout the twentieth century \cite{popper_open_1945, homans_bringing_1964, arrow_methodological_1994}, with methodological individualism taking on changing faces over time \cite{udehn_changing_2002}.

\paragraph{Social network analysis} Emerging in the mid-20th century, social network analysis (SNA) shifted focus to patterns of interaction among individuals, using sociograms to map relationships \cite{moreno_who_1934}. Inspired by early sociologists such as Simmel, SNA pioneers developed bipartite graphs with two types of nodes---individuals and the groups or events to which they belonged---to capture affiliation patterns \cite{simmel_conflict_1908, feld_focused_1981, mcpherson_hypernetwork_1982, breiger_duality_1974}. This duality enabled researchers to study overlapping memberships and community structure indirectly, through individual-level ties. However, despite tracking group affiliations, SNA largely retained MI’s emphasis on individuals as the primary units of analysis \cite{wellman_structural_1988, neal_duality_2023}. Groups appeared only as aggregations of pairwise ties or node attributes, without independent dynamics of their own.

As a result, bipartite models struggled to represent genuine group-level phenomena, such as the temporal dynamics of group memberships, relationships between groups, emergence of norms, or institutional memory. The assumption was that all group effects could be derived from individual interactions, missing essential feedback loops or emergent group states. In contrast, more recent higher-order network models, based on hypergraphs or simplicial complexes, allow direct modeling of group interactions as irreducible entities with nonlinear dynamics, providing tools that address some of the concerns of group realist perspectives. 

\paragraph*{Groups are real} In response to MI's growing influence, a number of scholars mounted a defense of groups as legitimate units of analysis \cite{campbell_common_1956, warriner_groups_1956, horowitz_concept_1953, wynne-edwards_animal_1962}. In \textit{Groups Are Real: A Reaffirmation}, sociologist Warriner challenges methodological individualism by framing the debate as one between nominalists---who view groups as mere aggregations of individuals (with all reality vested in the individual)---and realists, who regard groups as ontologically distinct entities. He also critiques interactionists, who try to balance individual and group perspectives but tend to default to individual-level explanations, since groups lack the clear boundaries and subjective experiences of individuals. But, just as individuals are not reducible to their biology, he insists that groups are not reducible to the individuals that compose them. 

At this point, scholars sought to ground group realism in observable mechanisms or formal models. Campbell (1956) argued that groups are real, but the organism-like analogy is misleading \cite{campbell_common_1956}. To make his argument, he drew from the \textit{common fate principle}---the Gestalt idea that elements moving together in unison share a degree of reality. This principle helps explain why groups are perceived as real even if they lack physical unity, like a body. Think of a flock of birds or a marching band; no physical boundary binds them, but their coordinated movement signals a degree of \textit{we-ness} (a property Campbell liked to call \textit{entitativity}). Meanwhile, Simon (1964) made the argument that organizational goals cannot be simply reduced to individual goals \cite{simon_concept_1964}. Instead, organizations operate through structured roles that shape behavior at every level. Simon argued that organizational goals are emergent; they do not simply reflect the preferences of CEOs, boards, or stakeholders but instead arise from the constraints imposed by institutional roles at every level of the organization \cite{simon_concept_1964}. Indeed, organizational goals may even conflict with those of individuals, and organizational survival can depend on how well it maintains institutions that preserve the alignment between individual and group goals. Simon's work bridges individualist and group-level modeling, showing how organizational goals arise from constraints on individual roles, not from the aggregation of preferences alone. 

Inspired by Simon's work, organizational scientists embraced group realism in that they favored small working groups as building blocks of organizations, rather than individuals. In particular, they see working groups as more ``real'' since they have explicit boundaries, differentiated roles, and exhibit specific functions within the organization \cite{hackman_design_1987}. This view is discussed at length in Leavitt's paper ``Suppose we took groups seriously...'' (1976), laying the foundation for a systems-level approach to small group performance, known today as team science \cite{leavitt_suppose_1974, hackman_design_1987, katzenbach_wisdom_1992, mathieu_evolution_2018, wuchty_increasing_2007, goodwin_science_2018, hall_science_2018}. Like Schumpeter before him, Leavitt sought to put aside metaphysical debates by framing the question as a methodological debate. Hence, he championed empirical research on how communication structures shape group outcomes---drawing on the experimental work of Bavelas and others \cite{bavelas_communication_1950, leavitt_effects_1951}. Since then, the irreducibility of group performance has become a mainstay in team science, yet we still lack a unified modeling framework informed by ideas from the metaphysics of groups. In other words, we need a modeling framework that can account for groups with varying temporal dynamics and capabilities. 

\paragraph*{The evolutionary dynamics of group-level features} Anthropologists and cultural evolutionists have emphasized the coevolution of human psychology and group dynamics to explain our unique capacity for large-scale cooperation \cite{boyd_cultural_1982, richerson_cultural_2016}. Human cognition is uniquely attuned to social learning---especially through conformity, prestige bias, and over-imitation---while deviant behavior is regulated through norms and reputational sanctions. A striking feature of human psychology is our tendency to over-imitate: children across cultures faithfully reproduce unnecessary steps in tasks \cite{nielsen_overimitation_2010, tomasello_what_2003}. Early over-imitation in children helps explain humans' unique propensity to learn and teach cultural behaviors later in life, from cooking to rituals to the internalization of corporate cultures. This motivation for specific actions and patterns of thought within groups, or norms, is further reinforced when contrasted with other groups at all ages.. From a modeling perspective, these learning biases promote greater within-groups similarity while preserving variation between them.

Such dynamics lend themselves to formal modeling via multilevel selection or cultural group selection. When traits benefit the group more than the individual, group-level selection can dominate---even if the trait is individually costly. As Darwin himself proposed, traits like patriotism or self-sacrifice---especially in the context of intergroup warfare---may evolve because groups that promote them outcompete more selfish ones \cite{darwin_descent_1871, bowles_did_2009}. Recent studies show that group extinctions due to violent intergroup competition occur on timescales of 500-1,000 years \cite{soltis_can_1995}. Crucially, this logic applies not only to tribes or ethnic groups but also to voluntary organizations such as churches, universities, and firms, where selection acts more rapidly and competition is often nonviolent \cite{henrich_weirdest_2020}.

In parallel, institutional theorists such as Nelson and Winter extended evolutionary models to firms, reframing Schumpeter's notion of ``creative destruction'' as a process of organizational selection. They argue that firms compete through varying strategies, technologies, and routines, some of which prove more adaptive than others \cite{nelson_schumpeterian_1982}. Unlike neoclassical economics, their approach emphasizes bounded rationality: agents operate with limited information and adopt heuristics, leading to persistent diversity rather than convergence to an optimum. In this context, concepts like path dependence were thought to illustrate how early choices---like the QWERTY keyboard---can lock in suboptimal solutions due to institutional inertia \cite{nelson_search_1993}. Once embedded, organizational routines become difficult to reverse. Douglass North extended this perspective by describing institutions as emergent systems that persist beyond individual lifespans through cultural transmission: ``learning embodied in individuals, groups, and societies'' that is ``cumulative through time'' \cite{north_rise_1973}. For modelers, these ideas support representing institutions as evolving entities in their own right, with group-level traits such as norms, routines, and decision structures that are not reducible to individual behavior.

Cultural evolution theory and new institutional theory developed research programs around group-level traits that can evolve, stabilize, and compete in ways that go beyond individual-level dynamics. With models, they sought to avoid the earlier ``Panglossian adaptationism''---the assumption that all observed traits are somewhat optimally selected \cite{campbell_how_1994, soltis_can_1995}. For modelers, the group-based approach raises two central challenges. First, group dynamics often cannot be captured solely through individual traits or pairwise interactions---even under assumptions of weak emergence. Second, group-level traits and individual incentives frequently diverge, especially in the context of intergroup competition or collective decision-making. 

When inquiring about the unique nature of human groups, authors have grappled with several recurrent ideas: the varying temporal dynamics of group behaviors, the role of inter-group interactions, the apparent irreducibility of groups to their members, and potential misalignment between group minds and agents. As with the shortcomings of bipartite graphs in SNA, these ideas hint at recurring dimensions of groups that would benefit from being formalized mechanistically. From smaller, sometimes fleeting groups who maintain public goods to organizational behemoths, all demand explicit modeling of group composition and cross-level feedback.

\section{Typology of groups and group-based modeling}\label{typology.intro}

The literature on group interactions is fragmented across communities, with ongoing disagreements about the underlying ontology of groups. Informed by the history of this literature, we define a typology of models of groups that integrates mathematical assumptions with the long-standing history of group interactions. 

\begin{itemize}
    \item \textbf{Persistence:} Are group interactions ephemeral or persistent? Do the interaction dynamics wash out any correlation between individual states?
    \item \textbf{Coupling:} How much do you need to know about non-members to predict group dynamics?
    \item \textbf{Irreducibility:} Can you predict a group's behavior based solely on its members, or do you need extra information about the group itself?
    \item \textbf{Misalignment:} Do groups behave in ways that reflect the preferences of their members?
\end{itemize}

\begin{SCfigure}
    \centering
    \includegraphics[width=0.6\linewidth]{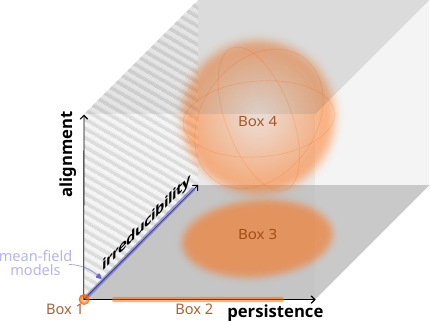}
    \caption{\textbf{A subset of the space of possible models of group interactions across three dimensions: persistence, alignment, irreducibility.} Simple mean-field models can capture irreducible group traits but do not capture the persistence of group members. As such, they cannot capture alignment of a group with its members; such as the adoption and rejection of group traits by group members over time. To expand on the spectrum of models, we present a simple mean-field model in Box 1, then add persistence in Box 2, irreducible group traits in Box 3, and explore the alignment dimension in Box 4.}
    \label{fig:cube}
\end{SCfigure}

\noindent We use this typology to classify broad families of models based on the kinds of higher-order interactions they can represent (see Fig.~\ref{fig:cube}). While this review focuses on models rather than the empirical aspects of human groups, we highlight connections to the empirical literature where possible. 

When discussing the existence of a group, persistence and coupling consistently emerge as key features. Early on, Simmel and others studied how group size and composition influence stability. They compared smaller, tightly knit groups based on intimate connections (primary groups) with larger groups that promote formal organizations, impersonal interactions, and specialized roles (secondary groups), such as bureaucracies \cite{cooley_social_1909}. Broadly speaking, the idea of group persistence is intertwined with that of interactions between groups, as groups are believed to engage in cooperation or competition. Strongly competing groups are more tightly bound, as they may restrict interactions to maintain their advantage. In contrast, cooperative groups are more open and can facilitate the mixing of people and ideas, leading to a reduction in group differences. Although permanence and isolation are key features, their meaning varies depending on the social groups in question. Both work teams and ethnolinguistic tribes are, to some extent, permanent or relatively isolated, but in notably different ways.

\subsection{Group persistence and intergroup coupling} \label{interface.section.persistence}

From a modeling perspective, we distinguish groups formed through momentary interactions and those persisting in time. We define momentary group interaction as a set of individuals engaged in some shared activity within a time window that is short compared to the characteristic timescale of the dynamics. In the physics of HONs, these interactions can be modeled as hyperedges within an annealed hypergraph (see \hyperref[tcolorbox:modelbox1]{Box 1}). Groups continuously form and dissolve, so the individuals with whom one interacts continuously change.

In contrast, group persistence generally induces dynamical correlations. For example, if an individual in a household gets sick, others in the household are likely to become infected sooner or later. Since the household persists over time---enforcing repeated interactions---the states of its members become correlated through the dynamics (here, a contagion process). We can therefore study how specific group interactions influence the local and global dynamics. In the limit of never-changing groups, these define a \textit{quenched} structure, representable as a static hypergraph, where hyperedges (groups) connect to each other through shared nodes (e.g., a household and a sports team with a member in common). In this case, neglecting dynamical correlations may lead to a poor understanding of the system's behavior. One can define models that, in addition to accounting for the quenched topology, are able to preserve dynamical correlations within groups by tracking the state evolution of each group \cite{burgio_network_2021}. 

The large amount of information used by such models makes them highly accurate but also computationally costly. Moreover, all that information is often difficult to access in real-world systems.

An alternative approach is offered by approximate master equations (AMEs) models \cite{hebert-dufresne_propagation_2010, st-onge_master_2021, burgio_characteristic_2025}. Although assuming an infinite, annealed structure, such models can work as satisfactory approximations for large quenched (or slowly varying) topologies due to the dynamical correlations they account for. Groups are assumed to be constantly reshuffling, but not uniformly at random; they rather do so while respecting the current probability distribution of group states (e.g. proportion of groups with $i$ active and $j$ inactive nodes, for every $i$ and $j$), so that dynamic correlations within groups are preserved. More sophisticated formulations of AMEs can also account for some dynamic correlations \textit{between} groups \cite{burgio_characteristic_2025}.


\begin{tcolorbox}[title= Box 1: Mathematical models of ephemeral groups, floatplacement=t]
\label{tcolorbox:modelbox1}
The continuous formation and dissolution of groups implies the absence of dynamic correlations between the states of individuals. Therefore, knowing the expected state of a random individual is all we need to describe ephemeral groups. Let us consider the simple case of binary-state dynamics, where agents are either `active' or `inactive', for which a single mean-field equation for the fraction $I$ of active individuals suffices. When both activation and deactivation are interaction-based processes, the equation reads   
\begin{equation}
    \frac{d}{dt}I = m\sum_{n=2}^\infty \frac{p_n}{n} \sum_{i=0}^n \binom{n}{i} I^{i} (1-I)^{n-i} \left[(n-i)\beta(n,i)- i\alpha(n,i)\right] \; .
\label{eq:ephemeral_1}
\end{equation}
Here, $m p_n/n$ is the expected number of randomly formed groups of size $n$ per individual, being $p_n$ the probability distribution of group sizes; $\binom{n}{i} I^{i} (1-I)^{n-i}$ is the probability that a group of size $n$ has $i$ active members, and $\beta(n,i)$ ($\alpha(n,i)$) is the rate at which inactive (active) individuals become active (inactive) due to interaction with peers, which is a generic function of $n$ and $i$. \newline

Another important class of dynamics, comprising basic contagion models such as susceptible-infected-susceptible (SIS) dynamics, is one where one of the two processes---say, deactivation---is a spontaneous, interaction-independent process. In this case, the dynamic equation becomes
\begin{equation}
    \frac{d}{dt}I = -\alpha I + m\sum_{n=2}^\infty \frac{p_n}{n}\sum_{i=0}^n \binom{n}{i} I^{i} (1-I)^{n-i} (n-i)\beta(n,i) \; .
\label{eq:ephemeral_2}
\end{equation}

The description given by Eqs.~(\ref{eq:ephemeral_1}) and  (\ref{eq:ephemeral_2}) is equivalent to an \textit{annealed} structure with an infinite number of individuals (nodes) and groups (hyperedges), where groups randomly form by uniformly sampling the population and are all reshuffled on a timescale much (ideally, infinitely) shorter than the timescale (here defined by either $1/\beta$ or $1/\alpha$) at which the dynamics unfolds. Both structural and dynamical correlations thus vanish in these models. \newline

\end{tcolorbox}


If groups persist over time, it then makes sense to ask if and how they interact with the outside social world. In \hyperref[tcolorbox:modelbox2]{Box 2}, coupled groups are those in which out-group states provide information about within-group dynamics. For example, children playing together can spread infections between households. Similarly, group coupling affects information flow in social contagions---polarized groups freely exchange information internally but not externally. In pairwise networks, modularity captures this, but perfectly modular networks differ from quenched, coupled ones; in the former, contagion remains a sum of independent influences. In contrast, higher-order structures introduce nonlinear reinforcement, where repeated interactions within persistent groups amplify contagion beyond independent pairwise transmissions.

The coupling is distinct from adaptive networks, where individuals can move between groups based on several mechanisms, such as leaving when dissatisfied \cite{gross_adaptive_2008, marceau_adaptive_2010,burgio_characteristic_2025} or ascribed migration \cite[ch.6]{mcelreath_mathematical_2007}. Instead, our modeling allows for recombination, where a mix of ephemerality with persistence of group dynamics can result in something new. By modeling how individuals form, reshuffle, or stay put, we can achieve an adaptive group-structured system. In some cases, this approach might be better suited to model group processes such as intergroup competition \cite{wilson_intergroup_2003, henrich_cultural_2004, richerson_cultural_2013}. In other cases, such as with social systems exhibiting strong polarization, coupling can instead be extended by adding a polarity to the coupling parameter. In this case, antagonistic influence could simply lead groups to adopt the opposite behaviors of what adverse groups are exhibiting, without requiring people to move \cite{smaldino_coupled_2021}.

The description of persistent group-based dynamics remains limited in that the state of a group is fully derived from that of its members. Group interactions in households impacting contagion dynamics seem reasonable enough, but what about clans within ethnolinguistic tribes competing with each other, or tacit organizational knowledge based on group-level cultural traits? To account for persistent cultural behaviors, we propose shifting our view from groups as sets of individuals engaged in multi-way interactions to a stronger notion of emergence based on group-level cultural traits \cite{smaldino_cultural_2014}.

\begin{tcolorbox}[title= Box 2: Mathematical models of persistent groups, floatplacement=t]
\label{tcolorbox:modelbox2}
To account for persistence, we must at least preserve correlations within groups, since the duration of a group interaction induces correlations among the states of its members. The latter can also change state because of their participation in other groups. A minimal model could use the approximate master equations. These describe the temporal evolution of the probability distribution $G_{n,i}$ that a group of size $n$ has $i$ active individuals (between 0 and $n$). The description tracks the processes within a focal group exactly and the processes within external groups in a mean-field fashion. The distribution $G_{n,i}$ thus evolves according to
\begin{align}
    \notag \frac{d}{dt}G_{n,i} =&~ (n-i+1)\left[\beta(n,i-1) + \rho_a\phi\right]G_{n,i-1}
    - (n-i)\left[\beta(n,i) + \rho_a\phi\right]G_{n,i} \\
    &+ (i+1)\left[\alpha(n,i+1) + \rho_d\psi\right]G_{n,i+1} - i\left[\alpha(n,i) + \rho_d\psi\right]G_{n,i} \; ,
    \label{eq:box2}
\end{align}
where
\begin{equation}
   \phi = \frac{\sum_{n,i}(n-i)\beta(n,i)G_{n,i}}{\sum_{n,i}(n-i)G_{n,i}} \; ,\ \
   \psi = \frac{\sum_{n,i}i\alpha(n,i)G_{n,i}}{\sum_{n,i}iG_{n,i}} \; ,
   \label{eq:box2_aux}
\end{equation}

are the respective probabilities of being activated or deactivated in random external groups. The factors $\rho_a$ and $\rho_d$ quantify the coupling between groups for the activation and deactivation mechanisms, respectively (notice that we absorbed the number of groups per agent in the definition of these couplings). Having independent couplings makes the model flexible to describe processes that are either interaction-based or spontaneous (in which case the respective coupling is put to zero; e.g., $\rho_d=0$ for SIS dynamics). The case of isolated groups corresponds to $\rho_a = \rho_d = 0$.

\end{tcolorbox}


\subsection{Irreducible group-level features} \label{interface.section.irreducible}

In many ways, the vast majority of group interactions are mediated by cultural group-level traits, as human behaviors are regulated by institutions. Consider how sports teams consist of players, staff, and management, all working together to optimize team performance. Teams succeed or fail not only due to synergistic individual performance but also due to persistent group-level traits, such as how management decides to allocate resources among its members \cite{turchin_ultrasociety_2016}. Cultural traits are considered to operate at the group level because they are shaped by the ratio of within-group and between-group competition, rather than by individual performance alone \footnote{This argument draws on the Price Equation, which shows that the ratio of within-group to between-group variance determines the conditions under which group selection can outweigh individual selection \cite{richerson_cultural_2016}. Here we avoid introducing yet another formalism to the paper and keep the text focused on the formalism discussed in the mathematical boxes.}. For example, if management chooses to distribute resources more evenly, it reduces within-group competition for the best contracts \cite{tiokhin_shifting_2024}. In this context, contracts promoting equity serve as one of many norms that facilitate cooperation among team members, contributing to the group's overall \textit{organization}. In \hyperref[tcolorbox:modelbox3]{Box 3}, this idea is captured by introducing an abstract group-level feature, $\ell$, which can be framed, for instance, as fostering altruistic behavior within teams. 

By assuming group-level independence, we can directly model how groups adopt a series of discrete levels of institutional strength (say $\ell = 0,\dots,L$) to enhance their capacity to promote individually costly behaviors. In the current form of the model, adopting stronger norms results in a faster adoption rate of costly behaviors among members, but this can be further elaborated (as we do in Sec. \ref{secion:alignment}). In our team example, modern teams could engage in increasingly sophisticated strategies to promote team equity. Just as coupling involves knowing the state of out-group members to predict the state of individuals within groups, irreducibility can be understood as the amount of information about the group required to predict the state of its members. When group state is perfectly correlated with that of its members, it is fully reducible. 

Without group-level traits, groups are indistinguishable if their members find themselves sharing the same states. Once we account for group states, these may vary or remain constant while shaping the dynamics within groups. For example, a team might change its communication platform, influencing group dynamics. But practically, we are interested in how those changes may lead to groups being more or less successful, exhibiting group-level fitness. In \hyperref[tcolorbox:modelbox3]{Box 3}, group-level fitness influences how groups decide to scale institutional strength up or down, based on a trade-off between achieving success and balancing other factors, such as the potential costs of strengthening group norms. With group-level features, we can begin to inquire about the impact of perceived success, or perhaps prestige, of neighboring groups on within-group dynamics \cite{richerson_cultural_2016, hebert-dufresne_source-sink_2022}.


\begin{tcolorbox}[title= Box 3: Mathematical models of persistent and irreducible groups, floatplacement=t]
\label{tcolorbox:modelbox3}
In irreducible groups, the dynamics affect both the state of the members of a group as well as group features $\ell$ (in the simplest setting being a number from $0$ to $L$) that do not directly follow from the state of the members. We therefore split the dynamics into two sets of transitions, one governing members and one governing emergent group features, like this 
\begin{equation}
    \frac{d}{dt}G_{n,i}^\ell = \frac{d}{dt}M_{n,i}^\ell + \frac{d}{dt}E_{n,i}^\ell \; .
\end{equation} \label{eq:box3} \\
The first set of transition rates, $\frac{d}{dt}M_{n,i}^\ell$, governs transitions related to the members of the group:
\begin{align}
\label{eq:box3_ind}
\notag \frac{d}{dt}M_{n,i}^\ell =&~ (n-i+1)\left[\gamma f_a(\ell) + \beta(n,i-1,\ell) + \rho_a \phi \right]G_{n,i-1}^\ell - (n-i)\left[ \gamma f_a(\ell) + \beta(n,i,\ell) + \rho_a \phi \right] G_{n,i}^\ell \\
    &+ (i+1) \left[ \gamma f_d(\ell) + \alpha(n,i+1,\ell) + \rho_d \psi \right] G_{n,i+1}^\ell - i\left[\gamma f_d(\ell) + \alpha(n,i,\ell) + \rho_d \psi \right] G_{n,i}^\ell \; .
\end{align}
As in \hyperref[tcolorbox:modelbox2]{Box 2}, upon adding a sum over $\ell$, $\phi$ and $\psi$ represent the influence of other groups on individual transitions. The function $f_a$ ($f_d$) provides the probability that an inactive/defective (active/cooperative) individual becomes active/cooperative (inactive/defective) due to the pressure exerted by the institution (encoded as a function its level $\ell$), and $\gamma$ is the rate at which such change of behavior has the chance to occur. To achieve desired outcomes, norms could also alter the way individuals interact, hence the potential dependence of the interaction rates on $\ell$. \newline

The second set of transition rates, $\frac{d}{dt}E_{n,i}^\ell$, governs the transitions for the group feature $\ell$. The latter might capture the level of group activity dedicated to promoting a certain behavior (e.g., the creation of group norms). We could then assign a perceived fitness $Z^\ell$ to level $\ell$ (potentially also a function of the state of its members, i.e., $Z^{\ell}_{n,i}$) and let group minds perform a biased random walk over the fitness landscape of group features $\ell$:
\begin{align}
\label{eq:box3_groups}
    \notag \frac{d}{dt}E_{n,i}^\ell =&~ h_{+}(n,i,\ell-1)\left(\mu +  \dfrac{Z^{\ell}}{Z^{\ell-1}}\right) G_{n,i}^{\ell-1} - h_{+}(n,i,\ell)\left(\mu + \dfrac{Z^{\ell+1}}{Z^{\ell}}\right) G_{n,i}^{\ell}  \\
    &+ h_{-}(n,i,\ell+1)\left(\mu +  \dfrac{Z^{\ell}}{Z^{\ell+1}}\right) G_{n,i}^{\ell+1} - h_{-}(n,i,\ell)\left(\mu + \dfrac{Z^{\ell-1}}{Z^{\ell}}\right) G_{n,i}^{\ell} \; .
\end{align}
Parameter $\mu$ weights unbiased versus fitness-driven exploration of institutions, while the rates $h_{\pm}(n,i,\ell)$ account for potential dependencies on the current state of the group (e.g., resources, costs) to scale institutions up or down.
\end{tcolorbox}

\vspace*{5mm}

We highlight some key findings based on the dynamics of models from \hyperref[tcolorbox:modelbox3]{Box 3}. Critical thresholds in terms of collective costs, benefits, and adoption rates determine the widespread adoption of costly behaviors, through either continuous \cite{st-onge_paradoxes_2024} or discontinuous \cite{hebert-dufresne_source-sink_2022} phase transitions. Moreover, patterns of ``institutional localization'', in which a specific institutional level dominates the fitness landscape within certain parameter ranges, are also observed \cite{st-onge_paradoxes_2024,hebert-dufresne_source-sink_2022}. This has raised questions about the conditions under which groups invest in scaling up their policies versus relying on other groups to bear the burden of increased institutional efforts while reaping the benefits within their own groups---what has been dubbed as ``institutional free-riding'' \cite{st-onge_paradoxes_2024}. Lastly, when individual behaviors are perceived as excessively costly (perhaps adopting equitable contracts by sports players in our example is seen as such), institutions may respond by intensifying their efforts to promote those behaviors \cite{st-onge_paradoxes_2024}. This ``call to action'' suggests that worst-case scenarios can drive institutional change, whereas more tolerable situations may be accepted, leading---counterintuitively---to lower overall adoption rates.

Irreducible features help distinguish between groups as mere containers of interactions and groups as entities with higher-level traits subject to selection. The group independence assumption is a useful simplification in that we can directly model institutions without having to explain their emergence. However, how institutions emerge and evolve is fundamentally connected to---among other aspects such as social hierarchy (see Section \ref{interface.section.conclusion})---how individuals perceive and respond to norms; in other words, to which extent group- and individual-level goals align to each other, a dimension we explore in the next section.

\subsection{Alignment of individuals and groups} \label{secion:alignment} 
Alignment arises when group behavior reinforces individual goals, and vice versa. In contrast, misalignment occurs when groups encourage or produce outcomes despite contradicting or neglecting the goals or preferences of their members. Alignment interacts with previous dimensions to capture a diversity of group phenomena in nature.

A weak form of (mis)alignment emerges already in reducible groups. For instance, alignment can manifest simply as flocks of birds or schools of fish as a result of individuals obeying local interaction rules, such as repulsion, attraction, or predator avoidance \cite{reynolds_flocks_1987, ioannou_predatory_2012}. These decentralized dynamics generate group-level benefits (e.g., anti-predator shielding) that align with individual fitness. 
Similar effects can arise in human systems.
Roger's Paradox, where over-reliance on social learning degrades the quality of decisions in changing environments \cite{rogers_does_1988}, is an example of emergent misalignment. When individuals copy others rather than directly engaging with the world, groups risk becoming unresponsive or locked into outdated norms \cite{torney_social_2015}. Condorcet’s jury paradox, where individually rational votes aggregate into irrational majorities, is another example. In both cases, individuals act sensibly based on local information, yet the group produces outcomes no one intended. Real-world examples abound: alarm calls become noisy under excessive imitation \cite{brown_social_2003}, and speculative bubbles form when investors herd without reevaluating fundamentals \cite{lo_wisdom_2022}. In all these cases, (mis)alignment arises within \textit{persistent but reducible} groups---structures where individual behavior correlates over time, but the group exerts no independent influence. 

Irreducible groups unlock a stronger form of (mis)alignment, as they can persist even when their goals diverge from those of their members. Consider researchers choosing between slower, more reproducible methods and faster, less rigorous ones. When success metrics reward publication volume, individuals are incentivized to cut corners---even if the group would benefit from quality-focused norms \cite{dawson_role_2022, tiokhin_shifting_2024}. This is not necessarily a case of bad actors engaging in p-hacking; rather, individuals may continue practices they have learned and seen rewarded, despite growing evidence of their detrimental effects \cite{smaldino_natural_2016}. Groups may attempt to enforce better norms through institutional policies, but unless those norms are strong, visible, and valued, individuals may simply defect through apathy. In this context, misalignment stems not from resistance but from indifference.

The divergence between individual and institutional levels is not just a product of faulty aggregation. Institutions evolve on different timescales and are often sustained by mechanisms like prestige \cite{way_gender_2016}, norm entrenchment \cite{richerson_cultural_2016}, or institutional lock-in \cite{nelson_search_1993}. They may persist even when they no longer serve their members, creating structural inertia that deepens misalignment. Conversely, institutions can promote better practices---such as transparency or methodological rigor---while individuals may exhibit inertia on shorter timescales, failing to adapt even when norms shift in that direction.

We explore the effect of alignment in irreducible groups more formally in \hyperref[tcolorbox:modelbox4]{Box 4}. This is done via a minimal but crucial modification of the model presented in \hyperref[tcolorbox:modelbox3]{Box 3}. Individuals no longer shifts their behavior passively under institutional forces, but do it actively based on their satisfaction with the public good. Institutions do not influence individuals' behavior directly, but only through the quality of the public good individuals perceive while those institutions are in place. Accordingly, individuals are more (less) likely to behave cooperatively---(mis)aligning with the institutional goal---when more (less) satisfied, independently of how strong the institutional pressure is. If \hyperref[tcolorbox:modelbox3]{Box 3} relies on a positive feedback loop where stronger (weaker) norms induce more (less) cooperative behavior, in turn leading institutions to scale up (down) through selection, the alignment dimension introduced in \hyperref[tcolorbox:modelbox4]{Box 4} can either radicalize or soften such loop, allowing new macro- and mesoscopic outcomes.


By treating institutions as persistent group-level states that both shape and respond to individual behavior, we can model the coevolutionary feedback between individuals and group structure. Misalignment, in this view, is not a statistical quirk; it is a dynamic process through which group norms and individual actions gradually diverge, potentially undermining collective goals. 

\vspace*{2mm}
\begin{tcolorbox}[title= {Box 4: Mathematical models of persistent, irreducible, and (mis)aligned groups}, floatplacement=t]
\label{tcolorbox:modelbox4}
Alignment in irreducible groups takes the same basic form as in \hyperref[tcolorbox:modelbox3]{Box 3} but with the additional component that individuals have opinions about how their institution manages their public goods. As a result, stronger norms do not necessarily translate to higher individual adoption rates. If those norms are not perceived as valuable by group members, these might ignore or even oppose them. On the contrary, satisfied members might further align to the institutions. The same holds for the interaction rates: individuals might spend more energy pushing their peers towards or against what norms promote depending on how much satisfied they are.
A minimal implementation of (mis)alignment makes the rates related to spontaneous transitions in Eq. (\ref{eq:box3_ind}) functions of, not only institutional strength $\ell$, but also of individuals' satisfaction with the public good. The probability $f_a(\ell)$ becomes $f_a(\ell, s(n,i,\ell))\equiv f_a(n,i,\ell)$ (analogously for $f_d$), where function $s$ quantifies satisfaction. The latter could be a group-averaged measure depending solely on $\ell$ (e.g., fitness $Z^\ell$) or one accounting for the specific context of a group (e.g., $Z^\ell_{n,i}$). Either way, we might assume it acts simply as a modulation of the transition rates.
\newline

The group-level set of transition rates, $\frac{d}{dt} E_{i}^\ell$, can be taken identical to Eq. (\ref{eq:box3_groups}). The overall dynamics allow us to explore how individual- and group-level traits coevolve driven by (mis)alignment.

\end{tcolorbox}

\section{Discussion}\label{interface.section.conclusion}

We outlined four dimensions of group models, integrating insights from higher-order interaction networks and the long-standing history of group interactions. We recognized the former to mainly address weakly emergent group behaviors, whether ephemeral or persistent. By incorporating group-level features, we improve our ability to represent the coevolution of individuals and institutions. 
Along the way, we proposed a hierarchy of models that can explore these dimensions within a nested family of dynamical systems. While these models can become mathematically complex, their critical properties remain solvable through a simple mathematical recipe (see Appendix).
Before concluding, we highlight some challenges and future directions in applying our framework to empirical research.

\paragraph*{Measuring group-level features} Measuring co-authorship is easier than measuring the norms and institutions shared by research groups. Individual interactions often seem easier to measure than group interactions. Yet, this does not mean that social networks are more real than group interactions. Individual interactions are not as easy to measure as they seem; social networks are subject to assumptions and are often error-prone \cite{young_robust_2020}. What we define as nodes and edges depends on the research question; for example, friendship can be defined reciprocally or not, leading to different networks \cite{butts_revisiting_2009}. It is also beneficial to remind ourselves that all social interactions occur within specific sociotechnical contexts. Individual productivity is easier to measure because we have made it so, and co-authorship is particularly prominent, as Western scientists have long attributed ideas to individuals \cite{henrich_weirdest_2020}. Ultimately, the key is to use the right tools for the work at hand. Researchers studying teams in science should model changing norms and policies, as the meaning of co-authorship varies over time and across different communities. The availability of group interactions does not exempt us from clear assumptions about their meaning.

That being said, how would we start to measure group-level features? Anthropologists, among other fields, made tremendous efforts to measure and characterize cultural groups throughout the world. In recent times, researchers have begun to address the challenges of codifying norms and policies, making them accessible through open databases \cite{slingerland_database_2024, turchin_seshat_2015, kirby_d-place_2016}. Another example is the Oxford COVID-19 Government Response Tracker data, which was a tremendous effort by public health researchers to track policies adopted by governments in response to the COVID-19 pandemic. A key part of their effort was ensuring intercoder reliability in how a team of experts quantified the strength of government responses. In general, this ability to organize scientific teams on a larger scale to agree on relevant norms has been shown to be key to modeling group-level features \cite{slingerland_coding_2020}. 

\paragraph*{Measuring misalignment} Strongly emergent misalignment is the most challenging dimension to estimate because it involves variability in individual preferences and group dynamics. How do we determine if a sports team's management or a research group is misaligned with the interests of its members? Worse still, what about misalignment in organizations such as a ``Mafia'' or a ``Church''? We must rely on evolutionary models; if the group-level feature is assumed to promote group success, in terms of higher fitness, then misalignment occurs whenever individual preferences are not aligned with that functional goal. Sometimes, it is obvious, as with suicide bombers willing to die for their group to prosper. Most of the time, this is more nuanced, as with religious acts that entail serious risks and significant investments of time, leading to increased performance of the group relative to other religious groups \cite{power_social_2017}. Modeling functional group behaviors has been controversial, but the combination of contemporary theoretical modeling and fieldwork can help us identify the necessary ingredients to determine when group-beneficial traits are favored, as one of many potential alternative stable states \cite{boyd_group_1990, boyd_culture_1988, boyd_group_2002,soltis_can_1995}. Once again, a key aspect of quantifying misalignment is intercoder reliability among domain experts when identifying latent norms and policies, as well as individual variability, which can then be used to make predictions. 

\paragraph*{Group minds redux} When early social scientists discussed the idea of ``group minds'', they made logical arguments about society exhibiting organism-like characteristics. This has led to a kind of Panglossian evolutionism, where group adaptation is everywhere while motivating a Victorian \textit{laissez-faire} \cite{campbell_how_1994}, which was later dismissed. In our typology, we are assuming that \textit{group-level features}, such as norms and institutions, do influence group-level fitness. Importantly, fitness is defined generally in our model, but it must be empirically specified. In the case of interfirm competition, for instance, empirical research continues to specify how exactly organizational breakthroughs in management and innovation improve overall firm success, whether through differential survival or proliferation \cite{richerson_cultural_2016}. It is related but not reducible to the idea of group minds as collective intelligence, or weak emergence. In this view, there is a similar ongoing effort to empirically assess how teams perform better than the sum of individual performance, most notably through complementary skills, communication strategies, or diversity of expertise \cite{page_diversity_2010}. As such, the question of whether particular group-level features and weak emergence entail the existence of group minds remains open and is subject to empirical inquiry, although it has been influenced by extensive modeling work.

It is worth mentioning that philosophers have argued at length in recent years about the meaning of groups exhibiting cognitive-like properties \cite{french_collective_1984, pettit_groups_2001}. As with collective intelligence, some aspects of it are generally accepted, such as ants exhibiting memory-like properties through pheromone trails \cite{gordon_ant_2010}. However, as with the organism-like analogy, what is meant exactly by a corporation lying about the fuel efficiency of its vehicles is still an ongoing ontological debate \cite{lackey_epistemology_2021}.
In its strongest form, philosophers argue that groups capable of believing, knowing, asserting, or even lying can be held accountable for their actions, effectively becoming moral entities---an issue that falls beyond the scope of this review \cite{lackey_epistemology_2021}. 

\paragraph*{Big groups and networked norms} Our typology is meant to represent group interactions, both small and large. However, some group interactions are too abstract and hierarchical to currently fit into our modeling work; for example, a federal government issuing an executive order like a mask mandate. As such, the key level of analysis for our typology is the level at which policies are implemented, for instance, local health departments. At this level, groups can be thought of as copying and learning from each other. There is also an additional layer of global top-down influence, referred to as ``higher-level'' political-economic institutions, which interact with both ``lower-level'' institutions, such as those related to kinship, marriage, religion, and people's cultural psychology. Only by developing models that can capture the joint influence of individual variability along with that of lower-level institutions can we begin exploring the larger, more diverse ``pluralistic'' political institutions \cite{henrich_big_2015, acemoglu_why_2013}.

Our family of models focused on models with binary states at the individual level and a single abstract group-level state that influences those states. Nothing prevents our work from being extended to more complex states at all levels. One more challenging generalization is to consider overlapping institutions and identity as part of the connection between groups. People have overlapping memberships, which are known to significantly impact the dynamics of systems. Similarly, norms and institutions often come in bundles, with specific networks of practices.

\paragraph*{Conclusion}  As the modeling of group dynamics and higher-order networks flourishes, we will continue to see new mathematical approaches to describe groups. By using our typology to compare models, we hope to not only clarify and classify existing efforts but also identify types of groups and group interactions that are currently ignored by these efforts.

\section{Acknowledgments}
The authors acknowledge support from the following sources: J.S-O \& J.L. from the Alfred P. Sloan Foundation (Grant \#G-2024-22498); R.H., G.B. \& J.L. from the US National Science Foundation (Grant \#2242829); G.B., T.M. \& L.H.-D. from the US National Science Foundation (Grant \#2419733); G.B. from the Ministry of Science, Innovation and Universities of the Government of Spain (Grant JDC2024-054148-I). Any opinions, findings, conclusions, or recommendations expressed in this material are those of the authors and do not necessarily reflect the views of the aforementioned financial supporters.

\bibliography{biblio}

\section*{Appendix}

\subsection*{Critical points of group-based models}

In building our typology of group-based models, we introduced a nested hierarchy of mathematical descriptions aimed at capturing different dimensions of groups. While our typology and main text are aimed at a wide audience, this appendix presents calculations for the specialized readers interested in the physics and dynamics of group-based models.
In the following, we show how the critical points of all models within our hierarchy can be computed through a common perturbative approach. 

In particular, we are interested in finding the invasion threshold separating the inactive absorbing phase, where every agent is inactive/susceptible/defective, from an active one where active/infected/cooperative agents are sustained. This is done by studying the impact of small perturbations around the inactive absorbing phase. More precisely, expressing the activation rate as $\beta(n,i) = \lambda \tilde{\beta}(n,i)$, we want to find the critical value $\lambda_\text{cr}$ such that, at equilibrium, the population is fully inactive for $\lambda < \lambda_\text{cr}$ and has non-zero activity for $\lambda > \lambda_\text{cr}$. Given the formal symmetry between activation $\beta(n,i)$ and deactivation $\alpha(n,i)$ rates in the models' equations, the calculation of the invasion threshold for deactivation (separating a fully active phase from an inactive one) follows through identical steps. We thus only present the derivation of the invasion threshold for activation, $\lambda_\text{cr}$.

For the model in \hyperref[tcolorbox:modelbox1]{Box 1}, a general closed form solution is accessible for $\lambda_\text{cr}$. The models in \hyperref[tcolorbox:modelbox2]{Box 2} to \hyperref[tcolorbox:modelbox4]{Box 4} allow for an implicit solution when deactivation (or activation) is an interaction-independent process (e.g., SIS dynamics). The recipe for obtaining the implicit solution is formally the same in all three cases: linearizing the equations, imposing stationarity, and finally relating the invasion threshold to the largest eigenvalue of a matrix associated with the obtained linear system by leveraging the Perron-Frobenius theorem.

\subsubsection*{\hyperref[tcolorbox:modelbox1]{Box 1}}

Let us first consider the case where both activation and deactivation are interaction-based processes. To ensure the inactive phase is absorbing, note $\beta(n,0) = 0$ must hold. Close to the inactive absorbing state, we can expand $dI/dt$ from Eq.~(\ref{eq:ephemeral_1}) around $I=0$ to find, at leading order,
\begin{equation}
    \frac{d}{dt}I(t) = m[\langle (n-1)\beta(n,1)\rangle - \langle \alpha(n,1)\rangle]I(t) \; ,
\end{equation}
where $\langle x\rangle$ denotes the average of $x$ over the group size distribution $p_n$. This is immediately solved to give at leading order,
\begin{equation}
    I(t) = \exp\{m[\langle (n-1)\beta(n,1)\rangle - \langle \alpha(n,1)\rangle]t\} \; .
\end{equation}
The small perturbation $I(t)$ will thus either grow or shrink exponentially depending on whether the argument of the exponential is positive or negative, respectively. Imposing that argument to be zero, using $\beta(n,i) = \lambda \tilde{\beta}(n,i)$, we thus find
\begin{equation}
    \lambda_\text{cr} = \frac{\langle \alpha(n,1)\rangle}{m\langle (n-1)\beta(n,1)\rangle} \; .
\end{equation}

\subsubsection*{\hyperref[tcolorbox:modelbox2]{Box 2}}

Assuming deactivation is an interaction-independent process, the dynamical equations take the form
\begin{equation}
    \frac{d}{dt}G_{n,i} = (n-i+1)\left[\beta(n,i-1) + \rho_a\phi\right]G_{n,i-1}
    - (n-i)\left[\beta(n,i) + \rho_a\phi\right]G_{n,i} + (i+1)\alpha G_{n,i+1} - i\alpha G_{n,i} \; ,
    \label{eq:box2_app}
\end{equation}
with $\phi$ given in Eq.~(\ref{eq:box2_aux}), for $n \in \{n_\text{min},\dots,n_\text{max}\}$ and $i \in \{0,\dots,n\}$.

We linearize Eqs.~(\ref{eq:box2_app}) around the inactive state $I=0$, being $I = \sum_{i,n} (i/n)G_{n,i}$ the fraction of active agents per group, via the expansion
\begin{equation}
    G_{n,i} = G_n\delta_{i,0} + g_{n,i} I + {\cal O}(I^2) \; ,
    \label{eq:G_expansion}
\end{equation}
where $G_n \equiv \sum_i G_{n,i}$ denotes the fraction of groups of size $n$. The variables $\{g_{n,i}\}$ must satisfy $\sum_{n,i}(i/n)g_{n,i} = 1$ from the definition of $I$, while $\phi = I \sum_{n,i} (n-i)\beta(n,i)g_{n,i}/\langle n\rangle + {\cal O}(I^2)$, assumed $\beta(n,0)=0$ $\forall n$ to guarantee the inactive state to be absorbing. To the lowest order in $I$, after conveniently dividing both sides by $I$, Eqs.~(\ref{eq:box2_app}) become
\begin{align}
    \frac{1}{I}\frac{d}{dt}G_{n,1} =&~ \rho_a\frac{n}{\langle n\rangle}\sum_{n,i} (n-i)\beta(n,i)g_{n,i} - (n-1)\beta(n,1)g_{n,1} + 2\alpha g_{n,2} - \alpha g_{n,1} \; ,
    \label{eq:box2_app_lin1} \\
    \frac{1}{I}\frac{d}{dt}G_{n,i > 1} =&~ (n-i+1)\beta(n,i-1)g_{n,i-1} - (n-i)\beta(n,i)g_{n,i} + (i+1)\alpha g_{n,i+1} - i\alpha g_{n,i} \; ,
    \label{eq:box2_app_lin2}
\end{align}
where we omitted the equation for $dG_{n,0}/dt$ $\forall n$ as only $n$ of the $n+1$ equations are linearly independent due to the conservation of the total probability $dG_n/dt = 0$.

Now, since right at the invasion threshold the small perturbation $I$ neither grows nor shrinks, we can impose $dG_{n,i}/dt = 0$ $\forall\{n,i\}$. In this way, we obtain a determined linear system of $M=\sum_{n=n_\text{min}}^{n_\text{max}} n$ unknowns, i.e., $\{g_{n,i}\}_{n,i>0}$. This can be immediately recast in the form of an eigenvalue problem,
\begin{equation}
    \vec{g} = A\cdot\vec{g} \; ,
\end{equation}
where $\vec{g} \equiv (g_{n_\text{min},1},\dots,g_{n_\text{min},n_\text{min}},\dots,g_{n,1},\dots,g_{n,n},\dots,g_{n_\text{max},1},\dots,g_{n_\text{max},n_\text{max}})$ and $A$ is a non-negative and irreducible $M\times M$ matrix. From the Perron-Frobenius theorem \cite{meyer_matrix_2000}, $A$ admits only one physical eigenvector $\vec{g^*}$ of real, non-negative entries and such eigenvector is associated to its largest eigenvalue, $\Lambda_\text{max}(A)$. Therefore, the critical condition for the invasion threshold is
\begin{equation}
    \Lambda_\text{max}(A) = 1 \; .
\end{equation}
Finally, using $\beta(n,i) = \lambda \tilde{\beta}(n,i)$, the invasion threshold $\lambda_\text{cr}$ is found as the smallest value of $\lambda$ for which $\Lambda_\text{max}(A) = 1$ admits a solution that satisfies $\sum_{n,i} (i/n) g_{n,i} = 1$. Notice that alternative derivations of an implicit solution for the invasion threshold for this model already exist \cite{hebert-dufresne_propagation_2010, st-onge_master_2021}.

\subsubsection*{\hyperref[tcolorbox:modelbox3]{Box 3} \& \hyperref[tcolorbox:modelbox4]{Box 4}}

Assuming deactivation is an interaction-independent process, the dynamical equations take the form
\begin{align}
    \notag \frac{d}{dt}G_{n,i}^{\ell} =&~ (n-i+1)\left[\gamma f_a(n,i-1,\ell) + \beta(n,i-1,\ell) + \rho_a\phi\right]G_{n,i-1}^{\ell}
    - (n-i)\left[\gamma f_a(n,i,\ell) + \beta(n,i,\ell) + \rho_a\phi\right]G_{n,i}^{\ell} \\
    \notag & + (i+1)[\gamma f_d(n,i+1,\ell) + \alpha(n,i+1,\ell)]G_{n,i+1}^{\ell} - i[\gamma f_d(n,i,\ell) + \alpha(n,i,\ell)]G_{n,i}^{\ell} \\
    & + \rho~\nu_+(n,i,\ell-1) G_{n,i}^{\ell-1} + \rho~\nu_-(n,i,\ell+1) G_{n,i}^{\ell+1} - \rho[\nu_-(n,i,\ell)+\nu_+(n,i,\ell)] G_{n,i}^{\ell} \; ,
    \label{eq:box4_app}
\end{align}
for $n \in \{n_\text{min},\dots,n_\text{max}\}$, $i \in \{0,\dots,n\}$, $\ell \in \{0,\dots,L\}$. The equation above generalizes the models in \hyperref[tcolorbox:modelbox3]{Box 3} and \hyperref[tcolorbox:modelbox4]{Box 4} by introducing the generic functions  $\nu_\pm(n,i,\ell)$ for the transition rate from trait $\ell$ to trait $\ell\pm 1$ due to inter-group interactions, assumed to satisfy $\nu_\pm(n,i,\ell) = \tilde{\nu}_\pm(n,i,\ell) + {\cal O}(I)$ as the fraction $I$ of active agents per group goes to zero. Equation (\ref{eq:box3_groups}) is recovered by setting $\nu_{\pm}(n,i,\ell) = h_{\pm}(n,i,\ell)\left(\mu + Z^{\ell\pm 1}/Z^{\ell}\right)$. Moreover, for \hyperref[tcolorbox:modelbox3]{Box 3}, $f_a(n,i,\ell)$ ($f_d(n,i,\ell)$) and $\alpha(n,i,\ell)$ reduce respectively to $f_a(\ell)$ ($f_d(\ell)$) and $\alpha$. 

We linearize around the \emph{free} state defined here as the one where every agent is inactive ($I=0$) and all groups have trait $\ell=0$ ($G_n^\ell\propto\delta_{\ell,0}$). To be absorbing, in addition to $\beta(n,0,\ell)=0$ $\forall \{n,\ell\}$, the free state requires $f_a(n,i,0) = 0$ and $\nu_+(n,0,\ell) = 0$ $\forall \{n,\ell\}$.

Generalizing what done in the previous section, Eqs.~(\ref{eq:box4_app}) are linearized around the free state through $G_{n,i}^{\ell} = G_n^{\ell} \delta_{i,0} + g_{n,i}^{\ell} I + {\cal O}(I^2)$ and $G_n^\ell = G_n\delta_{\ell,0} + q_n^\ell I + {\cal O}(I^2)$, that combined provide the expansion
\begin{equation}
    G_{n,i}^{\ell} = G_n\delta_{\ell,0} \delta_{i,0} + (q_n^\ell \delta_{i,0} + g_{n,i}^{\ell}) I + {\cal O}(I^2) \; ,
    \label{eq:G_expansion_3}
\end{equation}
being $G_n^{\ell} = \sum_i G_{n,i}^\ell$ the fraction of groups of size $n$ and trait $\ell$, and $G_n \equiv \sum_{\ell} G_{n}^\ell$ the proportion of groups of size $n$. The relations $\sum_{n,i,\ell} (i/n) g_{n,i}^{\ell} = 1$ and $\phi = I \sum_{n,i,\ell}(n-i)\beta(n,i,\ell)g_{n,i}^{\ell}/\langle n\rangle + {\cal O}(I^2)$ hold. To the lowest order in $I$, Eqs.~(\ref{eq:box4_app}) become
\begin{align}
    \frac{1}{I}\frac{d}{dt}G_{n,0}^{0} =& -\rho_a\frac{n}{\langle n\rangle} \sum_{n,i,\ell}(n-i)\beta(n,i,\ell)g_{n,i}^{\ell} + [\gamma f_d(n,1,0)+\alpha(n,1,0)]g_{n,1}^{0} + \rho\tilde{\nu}_-(n,0,1)(q_n^1 + g_{n,0}^{1}) \; ,
    \label{eq:box4_app_lin1} \\
    \notag \frac{1}{I}\frac{d}{dt}G_{n,1}^{0} =& ~\rho_a\frac{n}{\langle n\rangle} \sum_{n,i,\ell}(n-i)\beta(n,i,\ell)g_{n,i}^{\ell} - (n-1)\beta(n,1,0)g_{n,1}^{0} + 2[\gamma f_d(n,2,0)+\alpha(n,2,0)] g_{n,2}^{0} \\ &- [\gamma f_d(n,1,0)+\alpha(n,1,0)]g_{n,1}^{0}
     + \rho \tilde{\nu}_-(n,1,1)g_{n,1}^{1} - \rho \tilde{\nu}_+(n,1,0)]g_{n,1}^{0} \; ,
    \label{eq:box4_app_lin2} \\
    \notag \frac{1}{I}\frac{d}{dt}G_{n,0}^{\ell > 0} =& -n \gamma f_a(n,0,\ell)(q_n^\ell + g_{n,0}^{\ell}) + [\gamma f_d(n,1,\ell)+\alpha(n,1,\ell)]g_{n,1}^{\ell} \\
    &+ \rho \tilde{\nu}_-(n,0,\ell+1)(q_n^{\ell+1}+g_{n,0}^{\ell+1}) - \rho \tilde{\nu}_-(n,0,\ell)(q_n^{\ell}+g_{n,0}^{\ell})\; ,
    \label{eq:box4_app_lin3} \\
    \notag \frac{1}{I}\frac{d}{dt}G_{n,1}^{\ell > 0} =&~ n \gamma f_a(n,0,\ell)(q_n^\ell + g_{n,0}^{\ell}) - (n-1)[\gamma f_a(n,1,\ell)+\beta(n,1,\ell)]g_{n,1}^{\ell} + 2[\gamma f_d(n,2,\ell)+\alpha(n,2,\ell)]g_{n,2}^{\ell} \\
    \notag &- [\gamma f_d(n,1,\ell)+\alpha(n,1,\ell)]g_{n,1}^{\ell} +\rho \tilde{\nu}_+(n,1,\ell-1)g_{n,1}^{\ell-1} - \rho \tilde{\nu}_+(n,1,\ell)g_{n,1}^{\ell} \\
    &+ \rho \tilde{\nu}_-(n,1,\ell+1)g_{n,1}^{\ell+1} - \rho \tilde{\nu}_-(n,1,\ell)zg_{n,1}^{\ell} \; ,
    \label{eq:box4_app_lin4} \\
    \notag \frac{1}{I}\frac{d}{dt}G_{n,i>1}^{\ell} =&~ (n-i+1)[\gamma f_a(n,i-1,\ell)+\beta(n,i-1,\ell)]g_{n,i-1}^{\ell} -(n-i)[\gamma f_a(n,i,\ell)+\beta(n,i,\ell)]g_{n,i}^{\ell} \\
    \notag &+ (i+1)[\gamma f_d(n,i+1,\ell)+\alpha(n,i+1,\ell)] g_{n,i+1}^{\ell} - i[\gamma f_d(n,i,\ell)+\alpha(n,i,\ell)] g_{n,i}^{\ell} \\
    &+ \rho \tilde{\nu}_-(n,i,\ell+1)g_{n,i}^{\ell+1} - \rho \tilde{\nu}_-(n,i,\ell)g_{n,i}^{\ell} + \rho \tilde{\nu}_+(n,i,\ell-1)g_{n,i}^{\ell-1} - \rho \tilde{\nu}_+(n,i,\ell)g_{n,i}^{\ell} \; .
    \label{eq:box4_app_lin5}
\end{align}

As before, we impose stationarity at the invasion threshold, i.e., $dG_{n,i}^\ell/dt = 0$ $\forall \{n,i,\ell\}$. We first observe that $q_n^\ell$ and $g_{n,0}^{\ell}$ always appear summed together. Defining $\tilde{g}_{n,0}^{\ell} \equiv q_n^\ell + g_{n,0}^{\ell}$, we thus obtain a determined linear system of $M = (L+1)\sum_{n=n_\text{min}}^{n_\text{max}}(n+1)-\sum_{n=n_\text{min}}^{n_\text{max}}1 = (L+1)\sum_{n=n_\text{min}}^{n_\text{max}}n +\sum_{n=n_\text{min}}^{n_\text{max}}L$ independent equations (one constraint for each value taken by $n$ being probability conservation, i.e., $d(\sum_{i,\ell} G_{n,i}^\ell)/dt=0$) of $M$ unknowns, i.e., $\{g_{n,i}^{\ell}\}^{\ell=0,\dots,L}_{n=n_\text{min}\dots,n_\text{max};\ i=1,\dots,n}$ plus $\{\tilde{g}_{n,0}^{\ell}\}_{n=n_\text{min}\dots,n_\text{max}}^{\ell=1,\dots,L}$. We can finally recast the system in the form $\vec{g}=A \cdot \vec{g}$, being $\vec{g} \equiv (\dots,g_{n,1}^0,\dots,g_{n,n}^0,g_{n,1}^1,\dots,g_{n,n}^1,\dots,g_{n,1}^L,\dots,g_{n,n}^L,\dots,\tilde{g}_{n,0}^1,\dots,\tilde{g}_{n,0}^L,\dots)_{n=n_\text{min}}^{n_\text{max}}$ and $A$ an $M\times M$ non-negative and irreducible matrix. Applying the Perron-Frobenius theorem and using $\beta(n,i) = \lambda \tilde{\beta}(n,i)$, the invasion threshold $\lambda_\text{cr}$ is found as the smallest value of $\lambda$ for which $\Lambda_\text{max}(A) = 1$ has a solution that satisfies $\sum_{n,i,\ell} (i/n) g_{n,i}^{\ell} = 1$.

\end{document}